\documentclass[aps,prb,twocolumn,groupedaddress,amsmath,showpacs,10pt,amssymb]{revtex4}

\usepackage{graphicx}
\begin{document}
\title{Nonlinear shot noise in mesoscopic diffusive normal-superconducting 
systems}


\author{Markku P. V. Stenberg and Tero T. Heikkil\"a}
\affiliation{Materials Physics Laboratory, Helsinki University of Technology, 
P. O. Box 2200 (Technical Physics), FIN-02015 HUT, Finland}
\thanks{We acknowledge Ya. Blanter, J. C.
Cuevas, F. Hekking, A. Kolek, and F. Wilhelm for discussions and comments.
The numerical simulations have been performed at the Center for
Scientific Computing (CSC, Finland).
This work was supported by the Finnish Cultural Foundation and the Graduate 
School in Technical Physics at the Helsinki University of Technology.}
\begin{abstract}
We study differential shot noise in mesoscopic diffusive normal-superconducting
(NS) heterostructures at finite voltages where nonlinear effects due to 
the superconducting proximity effect arise. A numerical scattering-matrix 
approach is adopted. Through an NS contact, we observe that the shot noise
shows a reentrant dependence on voltage due to the superconducting
proximity effect but the differential Fano factor stays approximately
constant. Furthermore, we consider differential shot noise in the
structures where an insulating barrier is formed between normal and
superconducting regions and calculate the differential Fano factor as
a function of barrier height.
\end{abstract}


\maketitle



Shot noise is fluctuation of the current that is due to the discrete nature
of the charge carriers. It is the only source of noise 
at zero temperature. 
Current noise contains information on the physics of transport phenomenon
not contained in the conductance. 
A classical value $S=2e|I|=S_{\rm{Poisson}}$ for the noise, 
observed, e.g., in a vacuum diode,
is obtained in the tunneling limit, {i.e.},
when the transmission probabilities for open scattering
channels are small and there are no correlations among the charge carriers. 
However, two effects may reduce the noise below its Poissonian value: 
inelastic scattering (not considered in this paper) 
and reduced noise in channels with finite transmission 
amplitudes.~\cite{beenakker92} In a phase-coherent conductor, all the transfer 
coefficients are not necessarely small. Instead, e.g., in a diffusive
phase-coherent metallic wire the distribution function of the 
transmission coefficients has a bimodal form \cite{beenakker97} such that 
almost closed and almost open channels are preferred. 
This results in a shot noise that is
one-third of the Poissonian noise independently of the sample-specific 
properties such as the number of channels or the degree of disorder. 
\cite{beenakker92,nagaev92}

During the last decade the noise properties of mesoscopic conductors
have been under intense study (for a review, see
Ref.~\onlinecite{blanter2000}).
There has also been increasing interest
to comprehend the interplay of phase coherence and superconducting
proximity effect in mesoscopic physics.
Recently the doubling of the shot noise in normal-superconducting 
heterojunctions, predicted in the linear regime in Ref.~\onlinecite{jong94},
has been verified experimentally. \cite{jehl}
While the theory of classical shot noise and the quantum mechanical results in 
the linear regime have been discussed before, 
little attention has been paid to mesoscopic noise at finite voltages when the 
presence of superconductivity induces nonlinear behavior in the transport 
coefficients. 
In Ref.~\onlinecite{taddei2001} current correlations 
in very small hybrid NS structures at finite voltages were discussed. 
Eliminating the effects due to the finite size of the structure
and fully taking into account the effects arising in a diffusive phase 
coherent sample, however, requires larger structures to be studied.
In Ref.~\onlinecite{belzig2001} a counting-field approach to the 
Keldysh Green's-function
method was adopted to calculate numerically the statistics of current
in a normal wire connected to normal and superconducting reservoirs at finite
voltages. In this paper we use a well-established scattering-matrix approach
to calculate the differential shot noise at finite 
voltages in the presence of the proximity effect.

We find that in the presence of superconductivity, the differential shot 
noise follows the reentrance peak observed in conductance such that the 
differential Fano factor remains approximately constant. In the NS
structure the resulting differential Fano factor is twice its normal
value. The Fano factor can be roughly interpreted to be the ratio of the
effective charge-carrying unit and the unit charge.
Hence the doubling of the shot noise is a signature of Cooper-pair transport 
in the NS junction. 

In the second part of the paper, we consider 
differential shot noise in structures where an insulating tunneling barrier separates the normal and superconducting regions.
In the tunneling limit differential conductance essentially probes the
density of states in the superconducting side except that at zero
voltage and low temperature the reflectionless-tunneling effect
increases the differential conductance. Reflectionless tunneling arises 
because of the quantum coherence of electrons and Andreev-reflected holes 
that travel along the same paths in opposite directions scattering 
several times from the barrier and the disorder potential of the metal.
\cite{wees92,katalsky91,marmorkos93}
We calculate the differential shot noise as a function of voltage and
predict that the reflectionless-tunneling effect is present not only
in the differential conductance, but also in the differential shot
noise, such that the differential Fano factor stays constant.

We consider the zero-frequency shot-noise power in a two-lead 
system that is the $\omega=0$ limit of the Fourier-transformed current-current
correlation function
\begin{equation}
S=\int_{-\infty}^{\infty}dt
{\mathbf\langle}[\hat{I}(t)-\langle\hat{I}(t)\rangle][\hat{I}(0)
-\langle\hat{I}(0)\rangle]{\mathbf\rangle},
\end{equation}
where the current operator $\hat{I}(t)$ may be expressed through 
a scattering matrix $\mathbf{s}$ and $\langle\rangle$
denotes the quantum mechanical expectation value.
In the NS junction where one reservoir is normal and the
other is superconducting the two-terminal differential 
shot noise at zero temperature $T$ takes the form \cite{jong94} 
\begin{eqnarray}
\frac{1}{eG_0}\frac{dS}{dV} & = & 2\mathrm{Tr}\big[
\mathbf{s}_{11}^{ee}\mathbf{s}_{11}^{ee\dag}
(\mathbf{1}- \mathbf{s}_{11}^{ee}\mathbf{s}_{11}^{ee\dag})+
\mathbf{s}_{11}^{he}\mathbf{s}_{11}^{he\dag}
(\mathbf{1}-\mathbf{s}_{11}^{he}\mathbf{s}_{11}^{he\dag}) 
\nonumber \\
 &   & +2\mathbf{s}_{11}^{ee}\mathbf{s}_{11}^{ee\dag}
\mathbf{s}_{11}^{he}\mathbf{s}_{11}^{he\dag}\big].
\label{eq: sntot}
\end{eqnarray}
Here $\mathbf{s}_{11}^{he}$ ($\mathbf{s}_{11}^{ee}$) is the 
submatrix of the scattering matrix referring to the reflection 
as a hole (electron) of an electron incident in lead $1$, evaluated at 
$E=eV$. 
At $T=0$ the differential conductance of the NS junction 
at voltage V is given by 
\begin{equation}
G=G_0\mathrm{Tr}\left[\mathbf{s}_{21}^{ee}\mathbf{s}_{21}^{ee\dag}
+\mathbf{s}_{21}^{he}\mathbf{s}_{21}^{he\dag}
+2\mathbf{s}_{11}^{he}\mathbf{s}_{11}^{he\dag}\right],
\end{equation}
where the unit conductance of a single spin-degenerate channel is denoted by 
$G_0=2e^2/h$. 
In the absence of normal transmission for voltages below $\Delta/e$
Eq.~(\ref{eq: sntot}) reduces to
\begin{equation}
\frac{1}{eG_0}\frac{dS}{dV}=8\rm{Tr}\big[
\mathbf{s}_{11}^{he}\mathbf{s}_{11}^{he\dag}
(\mathbf{1}-\mathbf{s}_{11}^{he}\mathbf{s}_{11}^{he\dag})\big]
\end{equation}
and the conductance is directly proportional to the Andreev
reflection probability,
\begin{equation}
G_{\rm NS}=2G_{0}\rm{Tr}\big[\mathbf{s}_{11}^{he}\mathbf{s}_{11}^{he\dag}\big],
\end{equation}
the factor of two indicating the fact that Andreev reflection creates a Cooper pair 
in the superconductor.

\begin{figure}
\includegraphics[width=7cm]{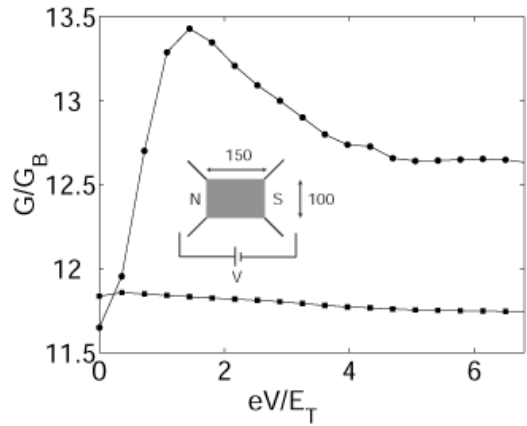}
\caption{Differential conductance as a function of voltage.
Circles, NS structure depicted in the inset; squares, N structure. 
The 95\% confidence interval for the relative error is $\pm 1\%$.
The results in Figs.~\ref{fig:gvse} - \ref{fig:fanovse}
were ensemble averaged over 600 realizations.}
\label{fig:gvse}
\end{figure}

We have studied two kinds of NS
heterostructures: a phase-coherent normal-metal wire
connected from the one end to a normal reservoir and from the other to a
long superconductor, and an NIS structure including
a tunneling barrier between the normal wire and the superconductor.
In our calculations we adopt a scattering-matrix approach and apply a numerical 
decimation method to truncate the Green's function of the 
two-dimensional structure. \cite{heikkila} 
The disordered normal-metal structure is modeled by a tight-binding 
Hamiltonian with the site energies varying at random within range 
$\left[-\frac{1}{2}w,\frac{1}{2}w\right]$.
Here we choose $w=\gamma$, where $\gamma$ is the nearest-neighbor 
coupling parameter. The calculated values of the observables are 
averaged over several impurity configurations.
The parameters of the structure are chosen such that the transport
through the normal metal is diffusive, i.e., 
the mean free path $l$ is much 
smaller than the length $L$ of the structure, which on the other hand is 
much smaller than the localization length $Nl$, where $N$ is the number of 
quantum channels ($l \ll L \ll Nl$).
The length scales of the structures in units of the lattice constant are
depicted in the insets of Figs.~\ref{fig:gvse} and \ref{fig:comparison}
illustrating the scattering geometries of the NS heterojunctions.

\begin{figure}
\includegraphics[width=7cm]{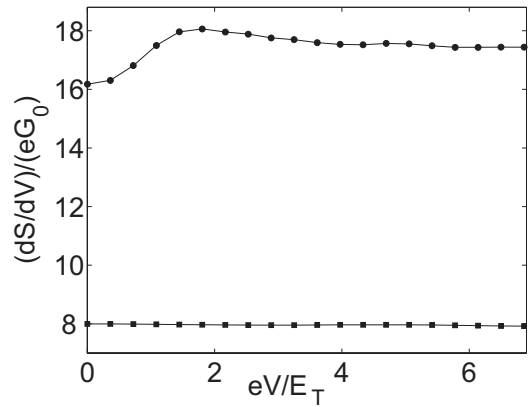}
\caption{Differential shot noise as a function of voltage. 
Circles, NS structure; squares, N structure.
The 95\% confidence interval for the relative error
is $\pm 1\%$.}
\label{fig:snvse}
\end{figure}

\begin{figure}
\includegraphics[width=7cm]{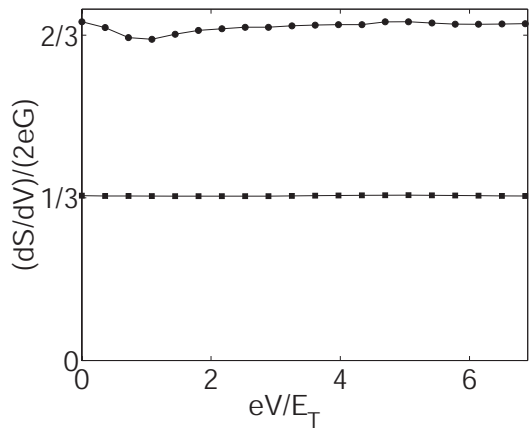}
\caption{Differential Fano factor $(dS/dV)/2eG$ as a 
function of voltage. 
The 95\% confidence interval for the relative error is
$\pm 1\%$.
Circles, NS structure; squares, N structure. The small variation in
the NS case is outside the error bars, but relatively smaller than the 
corresponding variations in $G$ and $dS/dV$.}
\label{fig:fanovse}
\end{figure}

\begin{figure}
\includegraphics[width=7cm]{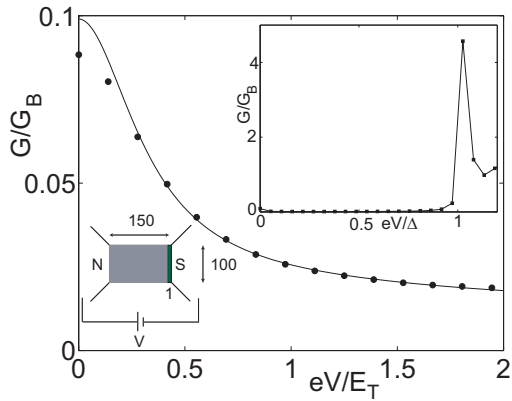}
\caption{Normalized differential conductance $G/G_B$ as a function of voltage
in the NIS structure depicted in the lower inset ($G_{\rm N}/G_B=10$). 
Circles, scattering-matrix approach; solid line, quasiclassical
theory \cite{qcselitys}. 
Upper inset: differential
conductance at the larger voltage scale. The 95$\%$ confidence 
interval for the relative
error is $\pm1\%$. The results in Figs.~4 - 6 were ensemble averaged over
300 realizations.}
\label{fig:comparison}
\end{figure}

The influence of the proximity effect on current noise is most conveniently 
seen in the differential Fano factor $(dS/dV)/2eG$.
In order to compare the electrical transport through normal 
and normal-superconducting structures we calculate 
$(dS/dV)/2eG$ in these two cases.
The calculated differential conductances for normal ($G_{\rm N}$) and 
NS ($G_{\rm NS}$) structures are plotted as functions of voltage
in Fig.~\ref{fig:gvse}.
As expected, in NS conductance we observe the well-known nonlinear 
reentrant behaviour due to the presence of the superconducting proximity 
effect, i.e, $G_{\rm NS}$ exhibits a maximum at energies of the 
order of a few Thouless energies $E_T=\hslash D/L^2$. 
At this energy scale $G_{\rm N}$ remains constant.
At zero voltage, normal shot noise is known to have a value
$S_{\rm N}=\frac{1}{3}S_{\rm{Poisson}}$. 
In the normal case there are no nonlinearities at this energy scale. 
Thus increasing voltage does not change this result 
and the differential
shot noise $d{S_{\rm N}}/dV$ for the normal structure remains constant as shown
in Fig.~\ref{fig:snvse}.
In the NS case, however, the differential shot noise $dS_{\rm NS}/dV$
exhibits a similar reentrant effect as $G_{\rm NS}$.
This is in agreement with the counting-field approach
of Ref.~\onlinecite{belzig2001}. 
Combining these two results, in the N structure, we observe that for 
the differential Fano factor the result 
$(dS_{\rm N}/dV)/2eG=1/3$ holds also for finite voltages, i.e., differential
shot noise has the value one third of the Poisson value. 
In the NS structure differential shot noise follows the reentrance peak 
observed in differential conductance, such that the 
differential Fano factor $(dS_{\rm N}/dV)/2eG \approx 2/3$ remains
approximately constant (Fig.~\ref{fig:fanovse}). Hence, the
differential Fano factor is twice the normal value reflecting the fact
that in an NS junction the current essentially results
from the uncorrelated transfer of Cooper pairs. 

\begin{figure}
\includegraphics[width=7cm]{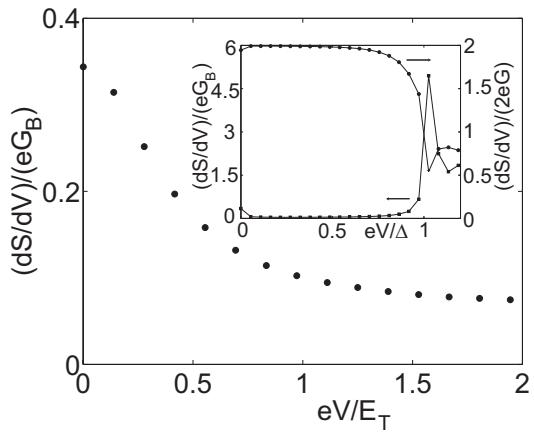}
\caption{Normalized differential shot noise $(dS/dV)/eG_B$
as a function of voltage. The 95$\%$ confidence interval for the
relative error is $\pm 1\%$.
Inset: Differential shot noise (squares)
and differential Fano factor (circles) at the larger voltage
scale.}
\label{fig:snrefl}
\end{figure}
\begin{figure}
\includegraphics[width=7cm]{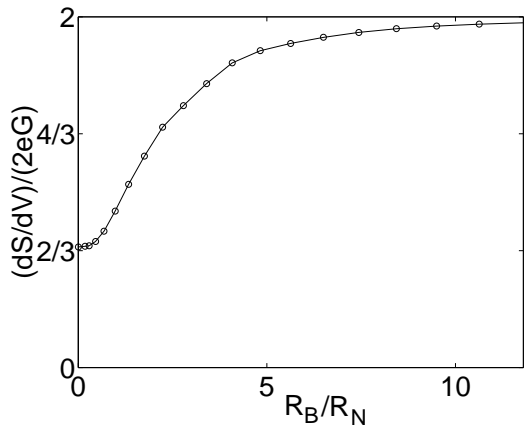}
\caption{Differential Fano factor $(dS/dV)/2eG$ in an NIS structure
at $V=0$ as a function of resistance of the tunneling barrier
$R_B/R_{\rm N}$ (measured in units of normal structure $R_{\rm N}$).
The 95\% confidence interval for the relative error is $\pm 1\%$.} 
\label{fig:fanovsrb}
\end{figure}
In Fig.~\ref{fig:gvse} we note that at zero voltage, $G_{\rm NS}$ 
is 2\% below $G_{\rm N}$.
This is due to a weak-localization effect resulting
from the quantum interference between time-reversed paths of the electrons
\cite{marmorkos93,nazarov95} yielding a different
contribution to $G_{\rm NS}$ than to $G_{\rm N}$. 

If the N wire is only weakly connected to the 
superconductor, the differential conductance
$G$ probes the density of states of the superconductor.
This is depicted in the upper inset of Fig.~\ref{fig:comparison}
where the conductance of a NIS structure is plotted as a function of
voltage. However, at zero voltage, the coherent interplay between the
Andreev-reflected and disorder-scattered electrons results in the
reflectionless-tunneling effect increasing $G_{\rm NIS}$ and creating a
peak around $V=0$ (Fig.~\ref{fig:comparison}).
The reflectionless-tunneling effect
may arise when there is a superconductor 
on the other side of the tunneling barrier and multiple scatterings by the 
tunneling barrier and by the disorder potential in the normal-metal area
take place. \cite{wees92,katalsky91,marmorkos93} 

As a test of our numerical results, we compare the obtained
differential conductance $G$ in the case of reflectionless tunneling to
the expressions derived from the quasiclassical theory of
nonequilibrium, inhomogeneous superconductivity
\cite{belzigetal,lambertraimondi} in the diffusive limit. To calculate
the current for the present setup, we need to solve for the transverse
distribution function (i.e., the symmetric part of the electron
distribution function around the chemical potential of the
superconductor) whose gradient determines the quasiparticle
current. In the limit $r_b=R_B/R_N \gg 1$, $eV \ll \Delta$, we obtain for
$G$ at $T=0$ (for details, see Ref.~\onlinecite{lambertraimondi} and the
references therein)
\begin{equation}
G=\frac{G_N(\sin 2\sqrt{v}+\sinh 2\sqrt{v})}{4 r_b^2
\sqrt{v}(\cos^2\sqrt{v}+\sinh^2\sqrt{v})
+\sin 2\sqrt{v}+\sinh 2\sqrt{v}}, 
\end{equation}
where $v\equiv eV/E_T$ and $G_{\rm N}$ is the normal-state conductance of
the diffusive wire. This is also plotted as a
function of $v$ in Fig.~\ref{fig:comparison} and agrees well with the
scattering-matrix approach. \cite{qcselitys} 

We have also calculated the differential shot noise 
as a function of voltage in the NIS structure (Fig.~\ref{fig:snrefl})
using the scattering-matrix method.
We observe that also the differential shot noise increases at zero voltage 
due to reflectionless tunneling.
The inset in Fig.~\ref{fig:snrefl} illustrates the differential shot noise 
and the differential Fano factor at larger voltages.
At voltages slightly above zero current and noise are suppressed since there 
are no  single-particle states in the superconductor and the Andreev reflection
probability is proportional to the square of the tunneling probability.
At $V=\Delta/e$ the differential shot noise follows the conductance peak. 
The interplay between the Cooper-pair and single-electron transport is most
clearly seen in the differential Fano factor. 
At voltages below $\Delta/e$ the Andreev reflection 
is the dominant charge transfer mechanism. Since the transmission
probability through the barrier is small, at low voltages we obtain 
a Fano factor which is twice the Poissonian value. 
As voltage approaches $\Delta/e$, the normal transmission probability 
increases and the system effectively behaves more like a normal 
conductor. Thus, the differential Fano factor quickly decreases to a
value near unity. In the inset of Fig.~5 the Fano factor at $V=0$ is
below the values obtained at somewhat higher voltages. This is due to
the fact that the chosen value for $r_b=10$ is not strictly in the
tunneling limit.

In order to illustrate the crossover from an ideal interface to the 
tunneling limit we have studied reflectionless tunneling by
calculating the Fano factor
as a function of the interface resistance $R_B$ at zero voltage. 
The resistance $R_N$ of the normal structure gives the characteristic 
scale for $R_B$, thus we plot the differential Fano factor as a function 
of $R_B$/$R_{\rm N}$ (Fig.~\ref{fig:fanovsrb}). At large values of $R_B$,
differential Fano factor approaches a limiting value two. 

In conclusion, we have studied differential shot noise in  
normal-superconducting mesoscopic structures in the nonlinear regime at 
finite voltages. 
The superconducting proximity effect manifests itself as a well-known 
nonlinear reentrance behaviour in the conductance at the voltages of the 
order of a few $E_T/e$. We have shown that also the shot noise exhibits 
a similar reentrance effect which keeps the differential Fano factor
approximately constant as a function of the voltage. In the second
part of the paper, we have considered a nonideal NS interface with an
insulating barrier between the normal and superconducting regions. We
find that also the differential shot noise exhibits a
reflectionless-tunneling effect observed as an enhancement of the
noise at zero voltage. Our calculations are consistent with the
quasiclassical results and other previous work.


\begin{thebibliography}{16}
\bibitem[{qcs()}]{qcselitys}
\bibinfo{note}{To obtain an agreement with the two approaches, we had to
  rescale the Thouless energy in Eq.(6) by a numerical factor of order 3.8.}
\end{thebibliography}
\end{document}